\batchmode
\documentstyle[12pt]{article}
\begin{document}
\topmargin=-.5in
\oddsidemargin=.1in
\evensidemargin=.1in
\vsize=23.5cm
\hsize=16cm
\def\thefootnote{\fnsymbol{footnote}}
\textheight=23.0cm
\textwidth=16cm
\baselineskip=24pt
\thispagestyle{empty}

\hfill{ITP-SB-97-60}\\
\smallskip
\hfill{October 1997}

\vspace{1.0in}

\setcounter{footnote}{0}

\centerline{\Large \bf HAIRS ON THE UNICORN}

\centerline{\Large  Fine structure of monopoles and other solitons\footnote{To
appear in Proceedings of the CRM-FIELDS-CAP Workshop ``Solitons'' at Queen's
University, Kingston, Ontario, July 1997 (Springer, New York 1998).}}
\vspace{.85in}

{\baselineskip=16pt
\centerline{\large Alfred S. Goldhaber}
\bigskip
\centerline{\it Institute for Theoretical Physics}
\centerline{\it State University of New York}
\centerline{\it Stony Brook, NY 11794-3840}}

\vspace{.5in}

\vspace{.5in}

\centerline{\Large Abstract}

\vspace{.5in}

Intrinsically stable or `fundamental' solitons may be decorated
with conserved charges which are pieces of those carried by
elementary particles in the same medium.  These `hairs' are always
significant in principle, and in the strong-coupling regime (where
solitons and particles exchange roles) may    
become major factors in dynamics.  

\vspace{1in}

\newpage

\noindent
{\bf I.  Monopoles and unicorns}

Long ago my developing fascination with magnetic monopoles led to an awareness of
a kinship between the monopole and another mythical beast, the unicorn.  As we
shall see shortly, the monopole (which was my personal entr\'ee into the subject
of solitons) occupies a very special place, as the only `fundamental'
soliton able to move freely in our (3+1)-dimensional world.

Before looking at the parallel between monopole and unicorn, let us go into 
another aspect of the monopole.  From Dirac's quantization 
argument\cite{DIRAC} we
know that the magnetic charge 
$g$ of a pole is quantized, and has a superstrong value.  Strong coupling is
incompatible with point structure of the coupled particle, and so the pole must
have internal geometrical dimension greater than its Compton wavelength by a 
factor
of order
$g^2$
\cite{GG}.  Because the natural quantum length scale is small compared to the size
of the object, this is a powerful indication that one should be able to describe
the pole as a classical field configuration, so that quantum consistency
conditions have brought us back to classical physics,
and hence to view the monopole as a soliton.  However, the quantum
consistency conditions for pure QED, or even $SU(2) \times U(1)$ electroweak
theory, still require quantized $g$, so clearly the charge cannot be spread out
over a length scale larger than the minimum for which the standard model is
accurate.  That is the basis for the claim below that existing or planned particle
accelerators won't have the energy to produce monopoles.  Now to the title:

The unicorn is a
wonderful creature.  It has many features in common with the monopole, including

\noindent
{\it 1.  Origin in medieval Europe}

The concept of a single pole, and the recognition that north and south magnetic
poles always come paired together, goes back at least to Peter the Pilgrim in
1269.  I don't know a first date for publication on unicorns, but it cannot have
been much later.  In both cases, one can find earlier but vaguer antecedents
in Asia.

\noindent
{\it 2.  Subject of a vast literature}

Until the last couple of decades, the literature on unicorns surely was larger
than that on monopoles.  Now the latest SLAC Spires listing contains 1125 titles
with the word `monopole', so the balance may have shifted.  In any case,
interest in both subjects remains strong.

\noindent
{\it 3.  Never confirmed or captured}

While there have been claims of finding either entity, they have never been
substantiated and accepted.  Nevertheless, the methods used both in seeking them
and in describing the results often have remarkable appeal and show great
insight.

\noindent
{\it 4.  Unique unity, not usual duplexity}

Like ordinary dipole magnets, fascinating and graceful two-horned animals are
commonplace, despite the fact that no known principle excludes the one-horned
possibility.

\noindent
{\it 5.  Illuminates much about the world}

By studying a physical theory to see if and how it accommodates monopoles, one
can learn a great deal about its structure, and about subtle consistency
conditions which otherwise might be overlooked.  In a similar way, contemplating
how people react to the concept of the unicorn teaches a great deal about human
nature, if not about animal biology.

\noindent
{\it 6.  Beautiful}

It is hard to imagine someone who could look at the Bayeaux tapestries and not
be entranced by the loveliness of the unicorn.  The perfect spherical
symmetry of the monopole, and the delicacy of its interaction with electrical
charges, has a charm which only seems to increase with acquaintance.

\noindent
{\it 7.  Poor cousins exist}

A monopole can be imitated in several ways.  Coulomb used the interaction
between nearby ends of two long magnets whose other ends were widely separated
to establish the inverse square law for magnetic poles with higher accuracy than
the corresponding law for electric charges.  Poincar\' e \cite{HP0} realized that
the interaction of an electric charge with a magnetic pole gives the same equation
of motion as for the purely mechanical system in which a rapidly spinning top is
kept aligned with the radial direction from some center to the location of a
massive particle.  For the unicorn, second-class imitations which come to mind
include the rhinoceros and the swordfish.

\noindent
{\it 8.  Instantly recognizable}

A monopole could be identified unambiguously by its ability to change the
quantized flux threading a superconducting loop.     For the unicorn, to see it
is to know it.  Of course, in both cases one must pay attention to effects which
might be fake indicators, but that is true in every aspect of life as well as
science.

\noindent
{\it 9.  Still hope of discovery}

For monopoles, we know from the success of the standard model that any
conceivable pole must have a mass greater than about 100 TeV, so they won't be
found at particle accelerators in the foreseeable future.  As theories exist
which are consistent with all observations at current energies but which have
monopoles at higher energies, they still could appear -- there might be a very
low flux of massive primordial monopoles detectable on earth.  For unicorns, the
possibility of finding one in some remote corner seems exactly that -- remote!
However, new breeding techniques with sophisticated genetic manipulation
conceivably might produce them.  Ironically, the best hope for monopoles lies
in observation, but for unicorns in experimentation!

\noindent
{\bf II. `Fundamental' and `complementary' solitons}

	Solitons have become an important part of physics, and yet their role 
remains a bit confusing.  At one level, things are quite simple:  a 
soliton is simply a massive object which can be described accurately as a 
classical nonlinear field configuration, perhaps absolutely stable or 
perhaps metastable.  This is an extremely attractive notion, because it 
may be the only alternative to introducing a particle as an elementary 
object, a quantum of some field.  With the soliton, the relation between 
particle and field suddenly develops a new possibility, where the 
particle is constructed from the field rather than vice versa.   
However, not all solitons are equal:  

Historically one of the first soliton-particles was the skyrmion -- a 
description of the nucleon as a map with winding number unity from 
ordinary space onto the three-dimensional surface of a sphere in four 
abstract Euclidean dimensions\cite {THRS}.
  Skyrme realized that this system could be quantized
either  with integer spin and isospin or half-integer spin and isospin.  The 
latter of course is what makes it a candidate to describe the nucleon.  
It also is an example of fractional soliton charge, as the soliton 
possesses isospin eigenvalues which are half-integer, unlike the integer 
values for the elementary pion field from which it is formally constructed.    
Further, it is a fermion where the pions are bosons.  Can you 
imagine a more stunning way to get something for nothing?  One 
might think scepticism about this `free lunch' explains why Skyrme got so little
response  when he first proposed the skyrmion, but the real reason seems to have 
been a near-total incomprehension of the concept 
that a particle could be described
as  a soliton at all.

One can only wonder what would have happened if there had been a more 
positive immediate reception to Skyrme's idea.  As we now know, it has an 
astonishing degree of success in describing the properties and 
interactions of nucleons with pions and even with each other.  
Conceivably such successes could have slowed down substantially the 
developments which led to QCD, now recognized as the underlying theory to 
which Skyrme dynamics can be a superb 
approximation in the domain of long-distance and low-energy 
interactions.  In this setting, it clearly is 
worthwhile to exploit the skyrmion for all it is worth, while still 
recognizing that ultimately it is a derivative from the fundamental 
theory.  In particular, the fractional charge and novel spin-statistics are 
not created by the Skyrme dynamics, only compatible with it, instead 
being based on the foundation of elementary quarks, with the peculiar 
charges and statistics resulting because in QCD the number of quark colors 
happens to be 
odd \cite{WF}.  Michael Mattis has given here a beautiful exposition of the subtle
but tight
connection between the Skyrme and QCD actions \cite{MPM}.  

This example of a very important `derivative' soliton naturally leads one
to ask ``Could	there be such a thing as a `fundamental' soliton, and 
what would that
mean?" At first this seems like a contradiction in terms, because the soliton is
described with a more or less elaborate structure expressed through field
configurations, so that it would appear in principle to be a composite.  However,
if the soliton has a uniqe type of charge, and if no conceivable variation of the
high-energy, short-distance dynamics of the fields used in its description could
destroy it, then one would be justified in calling it `fundamental'.  The
interesting point here is that only a few solitons can pass that test
\cite{me1}.  In one space dimension a `kink' has this property, meaning that some
field can minimize vacuum energy density for more than one value of the field, and
the kink interpolates between different allowed vacuum values at $x=\pm \infty$. 
In three space dimensions a magnetic monopole again passes the test, because the
Gauss law for magnetism assures that an isolated pole cannot be created or
destroyed.  Such a long-range field is the only known way to assure conservation
of a charge in more than one space dimension, and as electromagnetism is the
only field of that sort we have, the monopole really is the only option.
[An electrically-charged soliton would be stable if its electric charge were
not an integer multiple of the smallest elementary charge unit, but in any
theory which at least potentially could contain magnetic monopoles, this 
would be impossible to contrive.]   The
focus in the remainder of this discussion will be on fundamental solitons, and
since we  experience three space dimensions in our world, therefore especially on
monopoles. 

There is a common terminology which needs to be related to the terms introduced
above.  Solitons may be described as `topological' or `nontopological'.  The
fundamental solitons all are topological.  This is obvious for those in one
space dimension.  For a monopole, it follows because a Dirac string
\cite{DIRAC}   emanates from the pole and can only terminate on an antipole, or
(in different language) on every closed surface surrounding the pole the gauge
field corresponds to a nontrivial fiber bundle \cite{WY}.  On the other hand, as
discussed above, the skyrmion certainly is a topological soliton, even though it
is not fundamental.  Another important term is `duality':  Electric-magnetic
duality allows description in the weak-coupling domain where purely electric
particles are treated as elementary, and monopoles are treated as solitons,
while in the strong-coupling domain the roles of the particles and solitons are
interchanged.  From this point of view, `complementarity' is the best
term to describe the relationship of the Skyrme and QCD pictures.
The one gives a good first-order description of low-energy, long-distance
phenomena, while the other does the same for high energies and
short distances.  However, unlike charges and poles, the two types do not
coexist on either scale -- it is a matter of one or the other but not both.
A striking illustration of complementarity is the hybrid or `Cheshire
cat' model, where the inside of the nucleon is a chiral quark bag, and the
outside is a Skyrme field configuration \cite{rgb}.  There is no comparable
division for a fundamental soliton, emphasizing that the two types are
qualitatively different.

{\bf III.  Fractional and peculiar soliton charges}

Jackiw and Rebbi \cite {JR} observed that a soliton could carry charges which
are pieces of the charges carried by elementary particles.  We are
talking here about conserved, isolable charges, which have sharp eigenvalues
\cite{KS,R,GK,ASG}.  Evidently, only fundamental solitons could carry such
peculiar charges; otherwise, disappearance of a nonconserved soliton would leave
the charges with nowhere to go!  For the same reason, fundamental solitons and
their antisolitons must carry opposite values of the peculiar charges, modulo
the units found on an elementary particle. Even for fundamental
solitons, one would like to understand the mechanism by which the peculiar charges
are acquired, as it cannot be by simple binding of elementary particles to the
soliton.  What alternative is there?  Goldstone and Wilczek
\cite{GW} introduced a systematic way to study this question, by considering a
class of theories with an adjustable parameter, such that for some values of this
parameter there would be negligible coupling between the particle and the
soliton, and therefore no peculiar charge.  As the parameter changes slowly,
the
only way that the conserved charge can arrive is by adiabatic flow in from
infinity, and they found quantitatively what the flow would be for the
kinks and monopoles studied by Jackiw and Rebbi, verifying in particular
the latter's claim of half-integer fermion charge.  However, in the GW method 
there is no necessity to reproduce another feature of the original JR discussion,
namely a charge conjugation symmetry implying the existence of a
zero-energy fermion bound state in the presence of the soliton.   

Let us focus on the JR monopole example, which I tried repeatedly 
but unsuccessfully to
disprove, in the process nvertheless learning much which is valuable and
correct. Here let me emphasize what still makes this example mysterious,
and how one can at least partly penetrate the mystery.

First, let us review what they did.
JR \cite{JR} considered two examples of a
quantum Fermi field coupled to a specific classical, static Bose field
configuration (a  soliton).  In both cases the Dirac equation
for the
Fermi field
includes a single mode with zero frequency, and the rest of the spectrum is
completely symmetrical between positive and negative frequencies. 
Consequently, the state in which the zero-frequency mode is occupied and
that in which it is not ought to be charge conjugate to each other.  Since
they differ by one unit in fermion number $F$, they should be characterized
by
$F = \; \pm \frac{1}{2}$.  
The term `fermion number' is potentially confusing, as at least in 3+1
dimensions the number of spin-$\frac{1}{2}$ particles may possess only integer
eigenvalues.  However, one could imagine a new kind of `photon' coupled to all
fermions with a weight 1 and to all antifermions with a weight -1.  Then the
value of the `fermion charge' measured by this photon could in principle be
fractional for some special object which polarizes the fermion vacuum in a
particular way.  Thus, fermion charge coincides with fermion number for
collections of elementary fermions, but may have no simple relation to
fermion number for one of the special objects, the solitons.    

For the first JR example, where the Fermi
field is Yukawa-coupled to a sine-Gordon soliton in $1+1$ dimensions, the
treatment of Fermi
and Bose fields can be made symmetrical by bosonization of the Fermi
field. 
This makes it possible to account systematically for possible 
 back-reaction of the fermion on the boson degrees of freedom,
and confirms the suggestion that the object carries half-integer
$F$ \cite{SW}.  

For the monopole example, JR also noted
 a case where an extra label distinguishes two
zero modes, one corresponding to spin $+\frac{1}{2}$ in some direction 
and one to spin $-\frac{1}{2}$.  This is an example of 
charge \underline{dissociation} -- such a soliton
may
exist in
any of four nearly degenerate states, characterized by $F = \; \pm 1$, spin
$S = 0$
(spin singlet), or $F = 0, S = \frac{1}{2}$ (spin doublet).  Two charges 
(fermion
`number' and spin) which characterize a free fermion have been `torn apart'
so
that soliton states carry one 
charge or the other but not both.  Note that in this
example the states with $F=\pm 1$ should behave like bosons, in which case
those with 
$F=0$ would behave like fermions! 
Su, Schrieffer, and Heeger \cite{SSH} independently discovered 
charge dissociation in
their model of polyacetylene, with zero modes for two species of fermion, 
electrons
with spin up and electrons with spin down.

Recognizing that peculiar charge must be a vacuum polarization effect
allows us to make a link with a long-familiar phenomenon:
In  the presence of a medium, the charge of an elementary
excitation may be renormalized.  For example, an electron in an insulator with
dielectric  constant
$\epsilon$ carries a charge
$-e/\epsilon$, a fraction of the charge residing on that same electron in
vacuum.  One would not say that only a fraction of an electron is
present in the medium, just that the manifestation of the electron's presence
is different from what it would be in vacuum.

Recently it has been argued that the quasiparticles of the fractional
quantum Hall effect (FQHE) should be treated as electrons dressed by the FQHE
medium, so that their fractional charge is fractional in exactly the same sense as
for an electron inside an insulator \cite{GJ}.   There still is something
special about these quasiparticles, which can be described in two alternate ways:
1) The Aharonov-Bohm or Lorentz-force charge is renormalized by the same
factor as the local or Gauss-law charge \cite {GK}.  In ordinary insulators,
only the latter charge is renormalized.  2)  The electromagnetic field strength
is renormalized by the same factor as the Gauss-law charge \cite{GJ}.  This 
renormalization would be undefinable in ordinary three-dimensional systems,
but here refers to the ratio of
the effective field acting on excitations moving in the FQHE layer to the
field as measured in standard ways just outside the layer.  Whichever description 
one might prefer,
the quasiparticles
are  the unique 
light electrically charged
excitations in the FQHE medium, so that their charge can be considered
fractional only with respect to elementary excitations in \underline {other} 
media, such as the QED
vacuum.  


\noindent
{\bf  IV. Conditions for integer $F$}

Accepting that  fractional soliton charge is restricted to monopoles and kinks, 
are there any further limitations on fractional values in these cases? 
  To address this issue, at least with respect to fermion charge $F$
in situations with charge conjugation symmetry like that of Jackiw and Rebbi. 
let us
begin by recalling the

\noindent 
\underline{Theorem of Jackiw and Schrieffer}:  \cite{JS}  A
soliton whose spectrum is
invariant
under a charge conjugation symmetry $C$ which reverses the sign of $F$ may
have
integer
$F$, or half-integer $F$, but no other fractional value is allowed.

\noindent
\underline{Proof}:  For an isolated soliton, the only way $F$ (assumed
conserved) can
change is by scattering processes in which the number of fermions incident
differs
from the number emerging.  This means that two allowed values of $F$ must
differ by an integer.  $C$ symmetry implies that for every
allowed $F$, $-F$ also is allowed.  Therefore the difference $2F$ must be
an
integer, and $F$ must be either an integer or a half-integer.  If one
allowed
 $F$ is an integer (half-integer), so must be all the others, since
they differ by integers.  

        This theorem shows that to answer our question we need find
only what conditions must supplement charge conjugation symmetry
to
exclude half-integer eigenvalues. It also suggests a strategy for
determining those conditions.  If absorbing a fermion changes other
conserved charges of the soliton as well as $F$, then extra consistency
requirements may follow.  Therefore let us proceed by attending to other
charges carried by fermions.  The problem naturally divides fundamental
solitons into
two classes, (A) magnetic monopoles (interacting with fermions whose
electric charge has the minimum magnitude $e$ allowed by the Dirac
quantization
condition \cite{DIRAC}), and (B) all others.

        The reason for this division is that according to the
spin-statistics connection a fermion in $3 +1$ dimensions must carry
half-integer spin.  However, for a minimal electric charge in the field of
a magnetic monopole there is an extra electromagnetic angular momentum
which also is half-integer.  Consequently, it is possible for the
monopole to absorb a fermion with the appropriate magnitude of electric
charge without absorbing angular momentum.  For any
other kind of soliton, or for a monopole interacting with fermions whose
electric charges are even multiples of the minimum Dirac unit, absorbing a
fermion requires the absorption of a half-integer unit of angular momentum.
 Therefore spin is a suitable candidate for the extra quantum number
associated with depositing a fermion on a soliton for all cases in category
B.  For category A, the only apparent (and inevitably open) option for the
extra quantum number
is
electric charge.  Let us quote the results, leaving the rather long proof for
 case A to be presented
elsewhere \cite {me2}.

\noindent
\underline{Integer $F$ Theorem A}:  If the electromagnetic vacuum angle
$\theta$ vanishes, then a minimal-strength Dirac magnetic monopole symmetric under
fermion conjugation $C$ must carry integer $F$.

For all other solitons, including all condensed-matter
defects, we have

\noindent
\underline{Integer $F$ Theorem B}:  An object, other thn a minimal strength Dirac
magnetic monopole, symmetric under a
unitary charge conjugation symmetry $C$, with fermion number and spin both
sharp, must carry integer $F$.

\noindent
\underline{Proof}:   We are omitting the case of Theorem A, so that adding a
fermion to the object does change its spin by a half-integer.  $C$ is assumed to
commute with rotations, so that the unitarity  of $C$ implies its
commutation with angular momentum and spin.  Therefore charge conjugate
states must have the same spin.  This means that $2F$ must be an even
integer, as otherwise the states would differ in spin by a half-integer.
 Hence $F$ is an integer.  

The essence of the proof for Theorem A \cite {me2} is that, if $C$ reverses
electric charge
$Q$ as well as $F$, then $F=\frac{1}{2}$ implies $Q=\frac{1}{2}$. However,
fractional
$Q$
is equivalent to fractional vacuum angle \cite{WAMW}.  If $C$ does not reverse
$Q$, then the same reasoning as in Theorem B precludes fractional $F$.

Except for even-charge monopoles, all the candidate solitons for charge
$\frac{1}{2}$ under Theorem B must be kinks in one space dimension.  
Examples of these are discussed in \cite{ZKG,me2}. 

\noindent
{\bf V.  Questions about and applications of the Jackiw-Rebbi monopole}

In the semiclassical approximation studied by JR \cite{JR}, there are several
symmetries, in particular a charge conjugation symmetry $CP$ which reverses
both $F$ and $Q$, and a symmetry $G$ which only reverses $F$.  Neither of these
can be a simple symmetry if there is a single zero mode, because the two states
connected by the zero mode must differ in sign of the manifestly conserved
operator $U=e^{2\pi i T_a}$, where $a$ labels an arbitrary axis in isospin space
\cite{SW2}.  This does not invalidate the JR value for $F$, but draws attention
to a dilemma for the fully quantized  description.  If the mass of the fermion is
very small compared to that of the vector boson which determines the characteristic
radius of the monopole,  then the interaction of the fermion with the pole can be
described in terms of chiral `boundary conditions' on the Dirac wave functions at
the location of the pole, at least if electric Coulomb interactions are
negligible.  If there is to be exactly one zero-energy bound state, then is $u$ or
$d$  bound?  

	Seiberg and Witten \cite{SW2}, provide an important clue to solving this
mystery, once again by looking at a constraint associated with a very long
distance scale, as in the work of Goldstone and Wilczek \cite{GW}:  Suppose that
the fermion isodoublet has both an isoscalar and an isovector contribution to
its mass, such that (say) the $u$ fermion has neglible mass compared to $d$.  Then
the boundary conditions at the monopole  combine with requirements of energy
conservation to assure that, for small negative $u$ mass, $u$ but not $d$ has a
zero-energy bound state.  Further, as the $u$ mass goes from
slightly positive to slightly negative, there is an inevitable jump $\Delta \theta
= \frac{\pi}{2} mod \ \pi$ in the vacuum angle.  Thus, if that angle were zero
before, now all the conditions are obeyed for two states with $F=Q=\pm
\frac{1}{2}$.  Evidently, a similar jump in
$\theta$ should occur if instead the $d$ mass went through zero, and this time $d$
would have a zero mode.  Precisely at the point where the two masses are equal in
magnitude but opposite in sign, we have the ambiguous condition of the original JR
system.  It seems likely that if these masses are small compared to the
vector boson mass this point is a cusp, where the structure jumps, in such a way
that electric charge is exchanged between the bosonic dyon rotor degree of freedom
and the fermion vacuum polarization.  However, if the fermion and boson masses are
comparable, then the transition probably is smooth.  The difficulty of this
case comes because there is no large length scale which can be used to simplify
the problem. These issues are addressed more fully in \cite{me2}.
 
\noindent
{\bf VI.  Conclusions}

The fact that fundamental solitons indeed can carry pieces of the charges of
elementary particles is interesting, but
does it have wider implications?  At least one important positive indication is the
mutual reinforcement of JR fermion zero modes with the concepts of
electric-magnetic duality
 and supersymmetry
\cite {MO}.    
  Further, the monopole's fermion charge is
significant in the
strong-coupling domain where a monopole condensate may emerge 
to enforce both color confinement and chiral symmetry
breaking at the same time \cite{SW2}.  These developments are on top of 
results for condensed matter physics, illustrated in \cite{SSH,ZKG}.
Even though fundamental solitons  are
few, their role (and that of their peculiar charges) in the conceptual
understanding of physics is large.
 
\noindent
{\bf VII.  Acknowledgements}

Hidenaga Yamagishi, and more recently Frank Wilczek and Sidney Coleman helped me
learn about the vacuum angle and its important subtleties. Section II and
Ref.\cite{me1} were stimulated by Jeffrey Goldstone's scepticism that a skyrmion
with
$F=\frac{1}{2}$ could exist.  Michael Mattis corrected my early claim that
charge exchange between fermions and monopoles is suppressed regardless of fermion
mass.  This work was supported in part by the National Science Foundation.

\end{document}